\documentclass[twocolumn, amssymb, amsmath, showpacs, superscriptaddress, aps, prb]{revtex4}
\usepackage{graphics}
\usepackage{graphicx}
\usepackage{dcolumn} 
\usepackage{bm}
\usepackage{float}
\begin{document}

\title{Frustrated spin-1/2 square lattice in the layered perovskite PbVO$_3$}

\author{Alexander A. Tsirlin}
\affiliation{Max-Planck Institute for Chemical Physics of Solids, N\"othnitzer Str.\ 40, 01187 Dresden, Germany}
\affiliation{Department of Chemistry, Moscow State University, 119992 Moscow, Russia}
\author{Alexei A. Belik}
\affiliation{NIMS, Namiki 1-1, Tsukuba, Ibaraki 305-0044, Japan}
\author{Roman V. Shpanchenko}
\author{Evgeny V.~Antipov}
\affiliation{Department of Chemistry, Moscow State University, 119992 Moscow, Russia}
\author{Eiji Takayama-Muromachi}
\affiliation{NIMS, Namiki 1-1, Tsukuba, Ibaraki 305-0044, Japan}
\author{Helge Rosner}
\affiliation{Max-Planck Institute for Chemical Physics of Solids, N\"othnitzer Str.\ 40, 01187 Dresden, Germany}

\begin{abstract}
We report on the magnetic properties of the layered perovskite
PbVO$_3$. The results of magnetic susceptibility and specific heat
measurements as well as band structure calculations consistently suggest that
the $S=1/2$ square lattice of vanadium atoms in PbVO$_3$ is strongly
frustrated due to next-nearest-neighbor antiferromagnetic
interactions. The ratio of next-nearest-neighbor ($J_2$) to
nearest-neighbor ($J_1$) exchange integrals is estimated to be
$J_2/J_1\approx 0.2-0.4$. Thus, PbVO$_3$ is within or close to the
critical region of the $J_1-J_2$ frustrated square lattice. Supporting
this, no sign of long-range magnetic ordering was found down to 1.8 K.
\end{abstract}

\maketitle

The spin-1/2 square lattice provides a number of simple and important
models for theoretical physics. If one uses the Heisenberg Hamiltonian and
considers nearest-neighbor (NN) interactions only, a well-known
Heisenberg square lattice (HSL) is formed. This model has been
successfully applied to many transition-metal compounds (in
particular, undoped high-$T_c$ superconductors) and much of its
physics is now well understood.\cite{manousakis} A number of new
phenomena appear if next-nearest-neighbor (NNN) interactions are taken
into account. If both NN ($J_1$) and NNN ($J_2$) interactions are
antiferromagnetic, the spin lattice is frustrated since $J_1$ and
$J_2$ tend to establish different types of magnetic order. The
properties of the system are controlled by the value of
$\alpha=J_2/J_1$. If $\alpha$ is small, N\'eel order is favorable and
the limit of HSL is realized. If $\alpha$ is large, columnar
antiferromagnetic order is established. However, the main interest is
attracted to the intermediate region that lies close to the quantum
critical point at $\alpha_c=0.5$. The nature of the ground state in
this region is still disputed. Theoretical studies suggest different
spin-liquid scenarios (e.g.\ RVB ground state).\cite{sushkov,siurakshina2,chandra}

Most of the compounds realizing $S=1/2$ square lattice are well
described with HSL since NNN interactions are usually
negligible. Recently two new systems with $\alpha\gg 1$
(Li$_2$VOXO$_4$, X = Si, Ge) were a subject of extensive studies and
revealed columnar antiferromagnetic order.\cite{helge_prl,helge_prb,bombardi,melzi} However, no compounds in the critical region close to $\alpha_c=0.5$ have been reported so far.

Below we present magnetic properties of a novel compound, PbVO$_3$,\cite{roms,belik} that reveals $S=1/2$ frustrated square lattice (FSL). Both experimental and computational studies show that this
compound lies close to the critical region of FSL and does not undergo long-range magnetic ordering down to 1.8 K.

\begin{figure}
\includegraphics{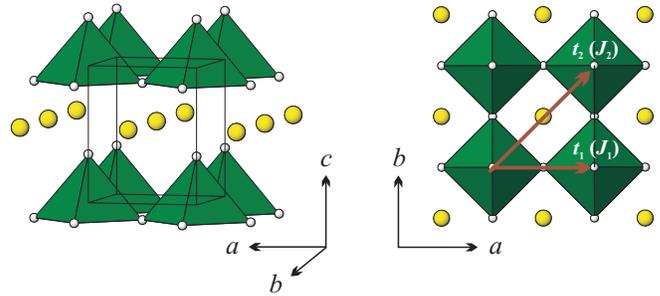}
\caption{\label{structure}(Color online) Perspective view (left panel) of the crystal
structure of PbVO$_3$ and projection along the [001] direction (right
panel). VO$_5$ square pyramids share common corners, lead atoms are
indicated by large spheres}
\end{figure}

PbVO$_3$ adopts a layered perovskite-type structure (space group
$P4mm$, $a=3.8001$ \r A, $c=4.6703$ \r A) shown in
Fig.\ \ref{structure}. This structure combines the absence of an
inversion center with a magnetic V$^{+4}$ cation, therefore PbVO$_3$
attracts considerable attention as a possible multiferroic.\cite{uratani, singh,ramesh1,ramesh2} Magnetic properties of PbVO$_3$ have not been reported so far.

Polycrystalline samples of PbVO$_3$ were prepared by a high-temperature
high-pressure technique in a belt-type apparatus. Stoichiometric
mixtures of PbO and VO$_2$ were placed into gold capsules and treated
at 950 $^\circ$C for 2 hours under a pressure of 5 GPa. The phase
composition of the prepared samples was checked by XRD.

The magnetic susceptibility of PbVO$_3$ was measured between 1.8 and 400 K in the fields $\mu_0H$ of 0.1, 1, and 5 T using a Quantum Design SQUID magnetometer. The specific heat was studied in the temperature range from 1.8 to 270 K with a commercial PPMS. 

\begin{figure}
  \includegraphics{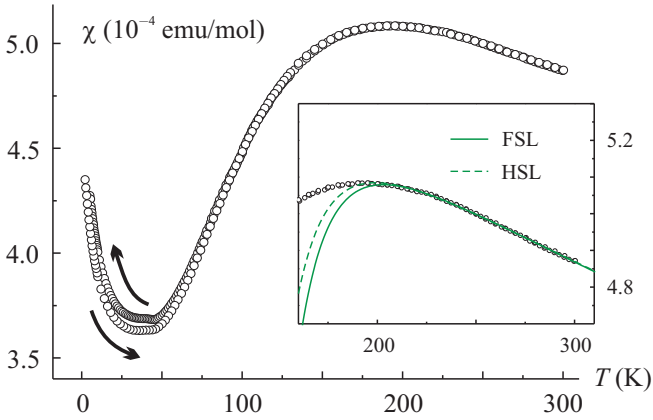}
  \caption{\label{suscept}(Color online) Magnetic susceptibility of PbVO$_3$ measured at 5 T, arrows denote the curves obtained at heating and cooling. The dashed and solid curves in the inset show the fits by high-temperature series expansion for HSL model and FSL model with $J_2/J_1=0.38$, respectively.}
\end{figure}

The study of the magnetic properties of PbVO$_3$ is complicated by the
presence of magnetic impurities in the samples under
investigation. All of the measured susceptibility curves revealed
anomalies at 90 and/or 350 K corresponding to PbV$_6$O$_{11}$ and
VO$_2$ respectively,\cite{foot1} although these impurities were not necessarily visible in XRD
patterns. Fig.~\ref{suscept} presents one of the best $\chi(T)$ curves
(no anomaly at 90 K) below 300 K, since the high-temperature part is
affected by VO$_2$. The features of this curve are typical for all
PbVO$_3$ samples within a variation of the temperature-independent
background from sample to sample due to different amount of
VO$_2$. These features are: i) a very broad maximum near $190-200$~K;
ii) the difference between the zfc and fc curves below about $50\,$K with
small humps near $43\,$K (or heating/cooling hysteresis visible at
Fig.~\ref{suscept}). One should note that the anomalies at $45-50$ K
may appear due to the influence of trapped oxygen that undergoes
condensation at about 50~K. These anomalies are especially strong if
the intrinsic signal is very small (as in the case of our
study). Nevertheless, the susceptibility data alone do not allow to
decide unambiguously whether the anomaly at 50~K is intrinsic to PbVO$_3$
(e.g. indicating a phase transition) or not. Therefore, we turn to
other experimental data in order to check the extrinsic nature of this
anomaly.

\begin{figure}
\includegraphics{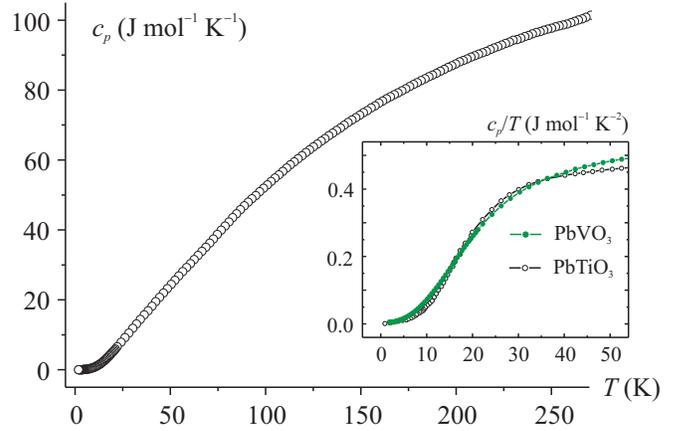}
\caption{\label{heat}(Color online) Specific heat of PbVO$_3$. The inset shows the
comparison with the diamagnetic reference compound PbTiO$_3$.}
\end{figure}

The specific heat curve (Fig.~\ref{heat}) is smooth between 1.8 and
270 K and suggests the absence of phase transitions in PbVO$_3$ in
this temperature range. The conclusion is supported by thermal
expansion\cite{belik} and resistivity\cite{roms,foot2} data. Finally, a neutron diffraction study at 1.5 K shows the absence of long-range magnetic ordering.\cite{roms} Thus, we conclude that the susceptibility anomaly at 50 K has extrinsic nature.

Square-pyramidal coordination of vanadium is known to give rise to
a non-degenerate $d_{xy}$ ground state for V$^{+4}$ (see Refs. \onlinecite{korotin,helge_prl}). The $d_{xy}$ orbitals lie parallel to the \mbox{V--O} layers, therefore one may expect that PbVO$_3$ is a 2D spin system. Indeed,
the broad maximum in the susceptibility curve (Fig.~\ref{suscept}) is typical for 2D spin systems while the upturn at low temperatures is usually ascribed to the paramagnetic contribution of
impurities and defects. Unfortunately most of the regions of the curve are unsuitable for fitting since the low-temperature part is affected by impurity contributions while the metal-insulator transition of VO$_2$ prevents us from using the data above 300~K.\cite{foot3} Thus, only the narrow region between the maximum ($T_{max}=190-200$ K) and 300 K is appropriate for the
fitting. Since the leading exchange integral $J$ is comparable to
$T_{max}$, high-temperature series expansions (HTSE) are applicable in
this region.

We use two types of HTSE corresponding to HSL and FSL models,
respectively.\cite{rushbrooke,helge_prb} A temperature-independent
contribution was added to the series as an adjustable parameter
($\chi_0$) in order to account for the contribution of VO$_2$ as well
as other (diamagnetic or van Vleck) contributions of this type. Both
models resulted in fits of similar quality with reasonable values of
$\chi_0\approx 2.5\cdot 10^{-4}$ emu/mol. We find $J=190$\ K for HSL
and $J_1=203$\ K, $\alpha=J_2/J_1=0.38$ for FSL. The $g$ value was fixed at
$g=2$ in order to get a more stable fit. Basically, we see that the
susceptibility curve is consistent with both scenarios -- HSL or FSL.

Specific heat data may provide additional quantitative information
about the spin system of PbVO$_3$. However, one has to estimate and
subtract the phonon part of the specific heat in order to reveal the
magnetic contribution. The leading exchange integral in PbVO$_3$ is
about 200~K therefore the magnetic contribution does not fall to zero
even at high temperatures and can not be separated from the phonon
part by a simple Debye-fit. A reference diamagnetic compound has to be
found in order to give a reliable estimate of the phonon contribution.

We tried to use PbTiO$_3$ as a non-magnetic reference. The crystal
structures of PbVO$_3$ and PbTiO$_3$ look similar but vanadium and
titanium adopt different coordination (square pyramid and octahedron,
respectively). This difference seems to be crucial: the $c_p(T)$
curves have crossings at low temperatures (see the inset of
Fig. \ref{heat}) indicating quite different phonon spectra. We are not
aware of any non-magnetic compound with layered perovskite-type
structure perfectly matching that of PbVO$_3$. Therefore, presently we
can not estimate the magnetic contribution to the specific heat of
PbVO$_3$.

Experimental data provide rather limited information about the
magnetic properties of PbVO$_3$, therefore we turn to computational
techniques. Band structure calculations are known as a useful tool for
estimating exchange integrals and understanding the properties of
low-dimensional spin systems. In particular, computational results
were helpful in the study of FSL compounds Li$_2$VOXO$_4$ and
provided the first realistic estimate\cite{helge_prl} \mbox{$\alpha\gg 1$} (supported by later neutron experiments, see Ref.~\onlinecite{bombardi}) in contrast to
the early experimental reports.\cite{melzi}

Scalar relativistic band-structure calculations were performed using
the full-potential local-orbital scheme (FPLO, version 5.00-19)
\cite{fplo} and the parametrization of Perdew and Wang for the
exchange and correlation potential.\cite{perdew} A $k$ mesh of 1152
points in the Brillouin zone (224 in the irreducible part) was used.

\begin{figure}
\includegraphics{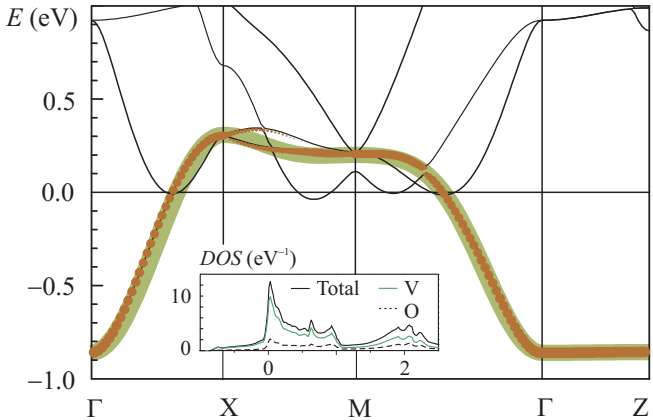}
\caption{\label{band}(Color online) LDA band structure of PbVO$_3$ near the Fermi
level. Dots show the contribution of V $3d_{xy}$ states, thick solid
line presents the fit of the tight-binding model. The inset shows the
density of states for PbVO$_3$ near the Fermi level.}
\end{figure}

We start with the LDA band structure (Fig. \ref{band}). The highest
occupied band reveals a significant contribution of V $d_{xy}$ orbital
consistent with the square-pyramidal coordination of V$^{+4}$ (see Refs. \onlinecite{korotin,helge_prl}). A simple one-band tight-binding model is fitted to this band. We find $t_1=0.132$ eV and $t_2=0.077$ eV for NN and NNN hoppings, respectively (see Fig. \ref{structure}). Long-range
in-layer hoppings do not exceed 0.004 eV and hence may be
neglected. The inter-layer hopping is also extremely small
($t_{\perp}=0.0012$ eV) suggesting a strongly two-dimensional spin
system in PbVO$_3$. The $t$ values are used to estimate
antiferromagnetic contributions to exchange integrals as
$J_i^{\text{AFM}}=4t_i^2/U_{\text{eff}}$ ($U_{\text{eff}}$ is an
effective on-site Coulomb repulsion). Assuming $U_{\text{eff}}=4$ eV
we find $J_1=203$ K, $J_{\perp}\approx 0.01$ K,
$\alpha=t_2^2/t_1^2=0.34$ in a perfect agreement with the FSL fit of
the susceptibility data. Note that the $\alpha$ value is a direct
result of the tight-binding fit and does not depend on
$U_{\text{eff}}$.

Thus, the tight-binding model supports the FSL scenario for
PbVO$_3$. However, this model assumes that all the interactions in the
system are antiferromagnetic. Sometimes it is not the case and one may
calculate total energies for different spin states in order to check
possible ferromagnetic contributions to the exchange integrals. LSDA
calculations for several simple spin states have been reported,\cite{uratani,singh} and we list these results in the first line of Table \ref{tab_fm}.

LSDA results for transition metal compounds are known to suffer from
an unreasonable neglect of correlation effects. Below we show that the
relative values of total energy are considerably changed if one takes
into account strong Coulomb correlation for vanadium $3d$ shell. The
most simple way to introduce such correlation within DFT is provided
by LSDA+$U$ technique.\cite{anisimov} We apply several physically
reasonable values of $U$ (further named $U_d$ in order to distinguish
it from $U_{\text{eff}}$) and fix $J=1$ eV.

\begin{table}
\caption{\label{tab_fm}The total energies of different spin states of
PbVO$_3$ (the values are given in meV relative to the energy of FM
state).}
\begin{ruledtabular}
\begin{tabular}{cccc}
  $U_d$ (eV) & A-AFM & C-AFM & G-AFM \\\hline
  0\ \cite{singh} & $+19.4$ & $-16.6$ & $-16.1$ \\
  4 & $-1.2$ & $-19.0$ & $-18.8$ \\
  5 & $-1.5$ & $-15.5$ & $-15.3$ \\
  6 & $-1.7$ & $-12.6$ & $-12.3$ \\
\end{tabular}
\end{ruledtabular}
\end{table}

Table~\ref{tab_fm} reveals quantitative rather than qualitative
dependence of the LSDA+$U$ results on the $U_d$ value. The energy
differences $(E_{\text{A-AFM}}-E_{\text{FM}})$ and
$(E_{\text{C-AFM}}-E_{\text{G-AFM}})$ corresponding to inter-layer
coupling $J_{\perp}$ are now comparable and small in contrast to the
LSDA results. The sign of the inter-layer interaction still is not
clear but the absolute value of $J_{\perp}$ has the order of units of
K. The in-layer interaction exceeds $J_{\perp}$ at least by an order
of magnitude. Thus, the spin system of PbVO$_3$ is two-dimensional
consistent with the crystal structure and the tight-binding results.

\begin{table}
\caption{\label{tab_j}$J_1$ and $J_2$ values calculated via total energies of states with different spin order.}
\begin{ruledtabular}
\begin{tabular}{cccc}
  $U_d$ (eV) & $J_1$ (K) & $J_2$ (K) & $\alpha=J_2/J_1$ \\\hline
  4 & 222 & 30 & 0.14 \\
  5 & 182 & 29 & 0.16 \\
  6 & 148 & 28 & 0.19 \\
\end{tabular}
\end{ruledtabular}
\end{table}

Now we use LSDA+$U$ to estimate in-layer interactions $J_1$ and
$J_2$. The results listed in Table \ref{tab_j} are in a reasonable
agreement with the experimental data and the tight-binding model,
although the $\alpha$ value is somewhat lower than
$0.34-0.38$. Nevertheless, LSDA+$U$ indicates considerable NNN
interaction. Thus, band structure calculations strongly support the
FSL rather than the HSL scenario.

Additional evidence for the frustration in PbVO$_3$ is found if one
considers the presence of long-range magnetic ordering in this
compound. The HSL tends to long-range order even at very weak
inter-layer coupling (for instance, N\'eel temperatures $T_{\text{N}}$ of undoped
high-$T_c$ superconductors have the order of hundreds of K, see Ref.~\onlinecite{highTc}). If $J_{\perp}$ in PbVO$_3$ is as small as 0.01 K (tight-binding result) $T_{\text{N}}\sim 0.2J_1\approx 40$ K.\cite{siurakshina} The phase transition at 40~K can hardly be missed
while analyzing experimental results. However, frustration effectively
reduces $T_{\text{N}}$ or even prevents the system from long-range magnetic
ordering at all.

Summarizing, all the experimental data available for bulk samples of
PbVO$_3$ do not provide evidence for long-range magnetic
ordering down to 1.8 K. Thermodynamic data (magnetic susceptibility,
specific heat, thermal expansion, and resistivity) do not show any
indications for intrinsic phase transitions, while low-temperature
neutron diffraction reveals the absence of long-range spin ordering at
1.5 K.

Note that the recently reported thin films of PbVO$_3$\cite{ramesh1,ramesh2} show a different magnetic behavior with
a possible magnetic ordering at 100--140 K. However, bulk phase and thin
film may have distinct properties due to slightly different structures
and strain effects. Detailed structural information for thin films of
PbVO$_3$ is not available, but the reported difference of the $c$
parameter of the unit cell (4.67 and 5.02 \r A in bulk and thin film,
respectively) suggests considerable change of the structure in thin
film as compared to bulk solid.\cite{foot4}

Thus, we conclude that magnetic frustration of the square lattice is a
crucial feature of PbVO$_3$. If $\alpha>\alpha_{c1}$
[$\alpha_{c1}=0.34$ (Ref. \onlinecite{sushkov}) or 0.24 (Ref. \onlinecite{siurakshina2})]
long-range ordering is suppressed and some kind of spin-liquid ground
state is formed. Unfortunately, we can not give a more precise estimate of
the $\alpha$ value for PbVO$_3$, hence we can not conclude if PbVO$_3$
falls into the spin-liquid region or is just close to its
boundary. Anyway, the frustration in PbVO$_3$ is strong enough to
suppress $T_{\text{N}}$ considerably (at least by an order of magnitude as
compared to the HSL estimate) or provide a disordered (spin-liquid or
spin-glass) ground state.

Summarizing, this study provides strong evidence that PbVO$_3$ is the
first example of $S=1/2$ square lattice system lying within or close
to the critical region of the FSL phase diagram and lacking for
long-range magnetic ordering down to low temperatures
($T_{\text{N}}/J<0.01$). A further study of the ground state and low-temperature
properties of this system is of high interest. Unfortunately, we have
to point out considerable difficulties in the synthesis of bulk
single-phase samples (and, moreover, single crystals) of PbVO$_3$. The
recent study of epitaxial thin films of PbVO$_3$
\cite{ramesh1,ramesh2} provides an alternative route for the
preparation of this interesting compound but thin films and bulk
solids may have strikingly different properties as we have mentioned
above.

In conclusion, we have shown that PbVO$_3$ reveals significant frustration of the square lattice that prevents this compound from long-range spin ordering down to 1.8 K. PbVO$_3$ lies very close to the critical region of the FSL phase diagram, and it may provide the first realization of disordered ground state for $S=1/2$ square lattice.
\begin{acknowledgments}
Financial support of RFBR (07-03-00890), GIF (I-811-257.14/03), Alfred Toepfer Foundation, and the Emmy Noether Program is acknowledged.
\end{acknowledgments}


\begin{thebibliography}{27}
\bibitem[{foo({\natexlab{a}})}]{foot1}
\bibinfo{note}{PbV$_6$O$_{11}$ undergoes ferromagnetic ordering at 90 K
  \cite{mentre} while VO$_2$ reveals a metal-insulator transition at about 350
  K. The latter transition is accompanied by a jump in the magnetic
  susceptibility (see, e.g. Ref. \onlinecite{vo2}).}
\bibitem[{foo({\natexlab{b}})}]{foot2}
\bibinfo{note}{The resistivity curve reported in Ref. \onlinecite{roms} shows an anomaly
  at 90 K. We assign this anomaly to ferromagnetic ordering of PbV$_6$O$_{11}$
  impurity, see Ref. \onlinecite{roms} for details.}
\bibitem[{foo({\natexlab{c}})}]{foot3}
\bibinfo{note}{The susceptibility of VO$_2$ is temperature-independent below
  300 K hence reasonable fitting of the $\chi(T)$ curve is possible.\cite{vo2}
  On the other hand, the contribution of VO$_2$ is temperature-dependent above
  330--340 K, and a proper correction for such contribution is impossible.}
\bibitem[{foo({\natexlab{d}})}]{foot4}
\bibinfo{note}{In particular, thin films may have slightly different local
  environment of vanadium, and such change will definitely influence magnetic
  interactions.}
\end{thebibliography}
\end{document}